\documentclass[a4paper,reqno,11pt,final]{amsart} \usepackage{array}
\usepackage[latin1]{inputenc} \usepackage{dsfont} \usepackage{ucs}
\usepackage{slashed} \usepackage{color}
\usepackage{amsmath,amsfonts,amssymb,amsthm,amstext,amscd,eucal}
\usepackage{bbold}
\usepackage{array}
\usepackage[T1]{fontenc}
\usepackage[latin1]{inputenc}
\usepackage{mathrsfs}
\usepackage[notref,notcite]{showkeys}
\usepackage{hyperref}

\hypersetup{%
  pdftitle = {Symmetric discrete
  quotients of supersymmetric Hpp-waves and spin structures},
  pdfkeywords = {supergravity, supersymmetry, spin structures, Hpp-waves},
  pdfauthor = {Hannu Rajaniemi},
  pdfcreator = {\LaTeX\ with package \flqq hyperref\frqq} }

\DeclareMathOperator{\AdS}{AdS}
 
\DeclareMathOperator{\SO}{SO} 
\newcommand{\Spin}{\mathrm{Spin}} \newcommand{\RR}{\mathds{R}}
\newcommand{\ZZ}{\mathds{Z}} 
 
 \newcommand{\dvol}{\mathrm{dvol}}
 
\newcommand{\diag}{\mathrm{diag}}
\newcommand{\Complex}{\mathds{C}}

\theoremstyle{plain}
\newtheorem{thm}{Theorem}
\newtheorem{case}{Case}
\newtheorem{conj}[thm]{Conjecture}

\declareslashed{}{/}{0}{.1}{\nabla} 
\DeclareMathOperator{\Tr}{Tr} 
 
\DeclareMathOperator{\Ker}{Ker}

\newcommand{\so}{\mathfrak{so}} 
 
 \newcommand{\Id}{\mathds{1}}
\newcommand{\half}{\tfrac{1}{2}}

\newcommand{\MUNCH}[1]{\relax}

\newcommand{\FIXME}[1]{\textcolor{green}{\textsf{#1}}}

\let\FIXME\MUNCH
\begin{document}
	
\title[Symmetric quotients of supersymmetric Hpp-waves]{Symmetric discrete quotients of
  supersymmetric Hpp-waves and spin structures}
	
\author[H.J. Rajaniemi]{Hannu J. Rajaniemi}
	
\address{School of Mathematics, University of Edinburgh, Scotland, UK}
\email{H.J.Rajaniemi@sms.ed.ac.uk} \date{\today}

\begin{abstract}
  We explore the importance of the choice of spin structure in determining
  the amount of supersymmetry preserved by a symmetric M-theory background constructed by
  quotienting a supersymmetric Hpp-wave with a discrete subgroup in the centraliser of its
  isometry group.
\end{abstract}
\maketitle
\tableofcontents
\section{Introduction} \label{sec:intro}
	
A (bosonic) M-theory background is usually defined as consisting of the data
$\left(M,g,F\right)$, where $\left(M,g\right)$ is a Lorentzian manifold with metric $g$
and $F$ is a closed four-form. In addition $F$ and $g$ obey certain field equations that
do not concern us here.

The authors of \cite{figueroa-ofarrill:_super_and_spin_struc} observed that it is possible
to construct examples of non-simply connected isometric M-theory backgrounds that have the
same geometry and four-form $F$ but admit different fractions of supersymmetry depending
on the choice of the spin structure. Therefore, it would seem necessary to include the
choice of spin structure in the data defining a M-theory background as well. The purpose
of this note is to illustrate this point further by considering backgrounds that are
Lorentzian symmetric spaces (Cahen-Wallach spaces), as opposed to the Freund-Rubin
solutions involving spherical space forms that were treated in
\cite{figueroa-ofarrill:_super_and_spin_struc}. We will show that at least for known
\emph{symmetric} M-theory backgrounds with with more than 16 Killing spinors the choice of
spin structure that preserves any supersymmetry appears to be unique. In particular, this
includes symmetric discrete quotients of M-theory pp-wave solutions.

The relationship between the choice of the spin structure of a Lorentzian symmetric space
and the dimension of the space of its \emph{twistor spinors} was first considered in
\cite{baum00}, and although the details are slightly different when we consider Killing
spinors, our techniques owe much to this previous work. In the case of non-conformally
flat symmetric spaces the twistor spinors actually agree with parallel spinors,
corresponding to supergravity Killing spinors when $F = 0$. Since this case was already
treated in detail in \cite{baum00}, we will focus on solutions that admit a nonzero
four-form flux. Orbifolds of 11-dimensional pp-wave solutions have also been considered
previously in \cite{michelson02}, but only for a very particular solution with 26
supersymmetries.

\section{Discrete quotients and spin structures} \label{sec:spin}

Let $\left(M,g\right)$ be a simply connected $n$-dimensional Lorentzian spin manifold. We
denote its isometry group by $I(M,g)$. Suppose $D\subset I\left(M,g\right)$ is a discrete,
orientation-preserving subgroup, and let $e_{B}$, $B = 0\ldots n-1$ be a
pseudo-orthonormal frame on $M$. Then for any $\gamma\in D$ at a point $x \in M$,
$d\gamma_{x}\in\SO(1,n-1)$ corresponds to the linear map that transforms $e_{A}(x)$ to
$e_{A}\left(\gamma(x)\right))$. There are now two possible lifts of $d\gamma_{x}$ to
$\Spin(1,n-1)$ since the covering map $\Spin(1,n-1)\rightarrow\SO(1,n-1)$ is two-to-one:
we denote these by $\pm\Gamma(x)$.

Now let $\mathcal{E}\left(D\right)$ be the set of all left actions of $D$ on
$M\times\Spin(1,n-1)$ satisfying
\begin{equation}
  \label{eq:action} \epsilon(\gamma)\left(x,a\right) =
  \left(\gamma(x),\epsilon(\gamma,x)\cdot a\right)~,
\end{equation} where $\epsilon(\gamma,x) = \pm\Gamma(x)$.

Elements of $\mathcal{E}\left(D\right)$ correspond to spin structures on $N = M/D$. The
spinor bundle corresponding to $\epsilon\in\mathcal{E}\left(D\right)$ is given by
\begin{equation}
  \label{eq:spinorbundle} S_{\epsilon} = \left(M\times
    \Delta_{1,n-1}\right)/\epsilon~,
\end{equation} Here $\Delta_{1,n-1}$ is the spinor module and
$\epsilon(\gamma)\left(x,\psi(x)\right) =
\left(\gamma(x),\epsilon(\gamma,x)\cdot\psi(x)\right)$. It follows that the
spinor fields $\psi$ on $N$ are the spinor fields on $M$ that satisfy
\begin{equation}
  \label{eq:spinorfields} \psi\left(\gamma(x)\right) =
  \epsilon(\gamma,x)\cdot\psi(x) ~.
\end{equation}
	
In particular, when $M$ is a M-theory background and $D$ also preserves the four-form $F$,
the Killing spinors on $N = M/D$ are the $\epsilon$-invariant Killing spinors of $M$.
	
\section{Hpp-waves in M-theory} \label{sec:cwm}

The M-theory Hpp-waves are supersymmetric solutions of eleven-dimensional supergravity
equipped with the metric of a Lorentzian symmetric space of the Cahen-Wallach type
\cite{FOPflux} and a null homogeneous four-form. In the light-cone coordinates
$x^{\pm},x^{i},i=1\ldots 9$ the Cahen-Wallach metric can be written as
\begin{equation}
  \label{eq:CWmetric} g =
  2dx^{+}dx^{-}+\sum_{i,j}^{}A_{ij}x^{i}x^{j}\left(dx^{-}\right)^{2} +
  \sum_{i}^{}\left(dx^{i}\right)^{2}
\end{equation} where $A_{ij}$ is a real $9\times 9$ symmetric matrix. If
$A_{ij}$ is nondegenerate, $(M,g)$ is indecomposable; otherwise it decomposes to
a product of a lower-dimensional indecomposable CW space and an Euclidean space.
If $A_{ij}$ is zero, \eqref{eq:CWmetric} is simply the metric on flat space. As
noted in the introduction, we will only be considering cases with nonzero $F$.

The moduli space of CW metrics agrees with the space of unordered eigenvalues of $A_{ij}$
up to a positive scale: this space is diffeomorphic to $S^{8}/\Sigma_{9}$, where
$\Sigma_{9}$ is the permutation group of nine objects\cite{FOPflux}. In particular, a
positive rescaling of $A_{ij}$ can always be absorbed by a coordinate transformation. It
is, of course, also possible to exhibit $(M,g)$ as a symmetric space by constructing its
transvection group $G_{A}$ for which \eqref{eq:CWmetric} is the invariant metric. We refer
the reader to \cite{FOPflux} for details.
	
A natural choice of $F$ is a four-form preserved by the symmetries of the CW metric that
also satisfies the field equations. As explained e.g. in \cite{simon05,FOPflux}, the
natural choice is a parallel form
\begin{equation*} F = dx^{-}\wedge\Theta~,
\end{equation*} where $\Theta$ is a $3$-form on $\mathds{R}^9$. The equations of
motion imply that $\Tr A = -\half |\Theta|^2$.
	
We will make use of a global pseudo-orthonormal frame:
\begin{eqnarray*}
  e_{+} & = &
  \partial_+,\\ e_{i} & = &
  \partial_{i}, \\ e_{-} & = &
  \partial_{-} - \sum_{i,j}^{}\half A_{ij}x^{i}x^{j}
  \partial_{+}
\end{eqnarray*} and the corresponding coframe
\begin{eqnarray*} \theta^{+} & = & dx^{+} + \half A_{ij}x^{i}x^{j}dx^{-}\\
  \theta^{-} & = & dx^{-}\\ \theta^{i} & = & dx^{i}~.
\end{eqnarray*}
For the purposes of this note, there is no need to distinguish between coordinate and
frame indices.
	
The only nonzero connection forms for the metric \eqref{eq:CWmetric} are
\begin{equation}
  \label{eq:connforms} \omega^{+i} = A_{~j}^{i}x^{j}dx^{-}~.
\end{equation}
The Clifford algebra convention we use is
\begin{equation}
  \label{eq:clconv}
  \Gamma_{A}\Gamma_{B} + \Gamma_{B}\Gamma_{A} = -2\eta_{AB}.
\end{equation}
where $\eta_{AB}$ is the flat Minkowski metric with $\eta_{-+} = 1$ and $\eta_{ij}=
-\delta_{ij}$. In other words, the flat metric has ``mostly minus'' signature. 
	
Supergravity Killing spinors are parallel sections of the supercovariant connection
\begin{equation*}
  \mathcal{D}_X = \nabla_X +\frac{1}{6}\imath_X F +
  \frac{1}{12}X^{\flat}\wedge F := \nabla_{X} + \Omega_{X}~,
\end{equation*}
where $X^{\flat}$ is the one-form metric dual of the vector field $X$ and the forms act on
spinors via Clifford multiplication. The spinor module is isomorphic to $\RR^{32}$ and
hence the maximum possible dimension for the space of Killing spinors is $32$. There are
actually two possible spinor modules: the one consistent with our conventions is the one
on which the action of the centre of the Clifford algebra is nontrivial. For convenience,
we denote the fraction of supersymmetry a M-theory background admits by $\nu =
\frac{1}{32}\dim\Ker \mathcal{D}$.
	
For the generic Hpp solution with arbitrary $A_{ij}$ and $\Theta$ one finds
(\cite{FOPflux}) that the solutions to the Killing spinor equation
\begin{equation}
  \label{eq:Killing}
  \mathcal{D}\chi = 0
\end{equation}
take the form
\begin{equation}
  \label{eq:KSeqn} \chi(\psi_{+}) =
  \exp\left(\frac{1}{24}\Theta_{ijk}\Gamma^{ijk}\right)\psi_{+}~,
\end{equation}
where $\psi_{+}\in\Ker\Gamma_{+}$ is a constant spinor. In other words, for the generic
Hpp solution, $\nu = \half$.

There is, however, a special point in the moduli space with
\begin{eqnarray}
  \label{eq:maxsusy}
  \Theta & = & \mu dx^1\wedge dx^2\wedge dx^3~,\\
  A_{ij} & =&
  \begin{cases} -\frac{1}{9}\mu^{2}\delta_{ij} & i,j = 1,2,3\\
    -\frac{1}{36}\mu^{2}\delta_{ij} & i,j = 4\ldots 9
  \end{cases}
\end{eqnarray}
which preserves \emph{all} supersymmetry. The explicit expression for the Killing spinors
of this background was given in \cite{FOPflux}:
\begin{multline}
  \label{eq:cwks} \varepsilon_{\psi_+,\psi_-}(x) =
  \left(\cos\left(\frac{\mu}{4}x^{-}\right)\Id -
    \sin\left(\frac{\mu}{4}x^{-}\right)I\right)\psi_{+}\\ +
  \left(\cos\left(\frac{\mu}{12}x^{-}\right)\Id -
    \sin\left(\frac{\mu}{12}x^{-}\right)I\right)\psi_{-}\\ -
  \frac{1}{6}\mu\left(\sum_{i\leq 3}^{}x^{i}\Gamma_{i} - \half\sum_{i\geq
      4}x^{i}\Gamma_{i}\right)\left(\sin\left(\frac{\mu}{12}x^{-}\right)\Id -
    \cos\left(\frac{\mu}{12}x^{-}\right)I\right)\Gamma_{+}\psi_{-}~,
\end{multline}
where $I = \Gamma_{123}$, $I^2 = -\Id$ and $\psi_{\pm}\in\Ker\Gamma_{\pm}$ are arbitrary
constant spinors.

In addition to the maximally supersymmetric Hpp-solution and the generic $\half$-BPS
solution with arbitrary $A$, there are a number of other interesting loci in the Hpp-wave
moduli space. In \cite{ChrisJerome} Gauntlett and Hull constructed Hpp-solutions admitting
``exotic'' values of $\nu = \frac{9}{16},\frac{5}{8},\frac{11}{16},\frac{3}{4}$. These
solutions possess Killing spinors that lie in $\Ker\Gamma_{-}$(often referred to as
supernumerary Killing spinors) in addition to the ``generic'' ones given in
\eqref{eq:KSeqn}. As we will see, in these cases the 3-form $\Theta$ takes a very
particular form.

We briefly outline how one obtains the form of the Killing spinors in these cases. Since
$\Omega_{i}\Omega_{j} = 0$ for all $i,j$, a Killing spinor $\varepsilon$ can always be
written in the form
\begin{equation}
  \label{eq:exoks}
  \varepsilon_{\psi_{+},\psi_{-}}(x) = \left(\Id +
    x^{i}\Omega_{i}\right)\chi~,
\end{equation}
where $\Omega_{i} = -\tfrac{1}{24}\left(\Gamma_{i}\Theta +
  3\Theta\Gamma_{i}\right)\Gamma_{+}$ and
\begin{eqnarray*}
  \chi_{+} & = &
  \exp\left(-\frac{1}{4}x^{-}\Theta\right)\psi_{+}~,\\ \chi_{-} & = &
  \exp\left(-\frac{1}{12}x^{-}\Theta\right)\psi_{-}~.
\end{eqnarray*}
As in the generic case, $\psi_{+}$ is an arbitrary constant spinor annihilated by
$\Gamma_{+}$. However, now the $\psi_{-}\in\Ker_{-}$ are not arbitrary. Since $\Omega_{i}$
always involves $\Gamma_{+}$, substituting \eqref{eq:exoks} into the Killing spinor
equation \eqref{eq:Killing} imposes no extra conditions on $\chi_{+}$. But further
analysis reveals that \eqref{eq:Killing} can be split into independent $x^{-}$ - and
$x^{i}$ -dependent parts, the former of which gives the form of $\chi_{-}$ and the latter
can be written as
\begin{equation}
  \label{eq:chiminus} \left(-144\sum_{j}^{} A_{jk}\Gamma_{k} +
    X^{2}_{j}\Gamma_{j}\right)\chi_{-} = 0~,
\end{equation}
for each $j$, where $X_{j} = \Gamma_{j}\Theta\Gamma_{j} + 3\Theta$. Finding solutions with
supernumerary Killing spinors amounts to finding solutions to this equation. 

Let us assume that $A$ has been brought to a diagonal form via an orthogonal
transformation so that $A = \diag(\mu_1,\mu_2\ldots \mu_9), ~\mu_{i}\in\RR$. Then in
order to find solutions to equation \eqref{eq:chiminus}, we must ensure that the action of
$X_{j}^2$ on spinors is diagonalisable. Since $X_i$ involves the 3-form $\Theta$, the
natural next step is to find a diagonalisable ansatz for $\Theta$. To do this, we actually
need to complexify the spinors, i.e. look at the action of $\Theta$ on the bundle
$S \otimes\Complex$. \FIXME{What is the proper way to do this? I think this is okay, since in
  every case we will actually look at the action of $\Theta^2$ terms.}

It is well known that the Lie algebra of $\SO(16)$ is isomorphic to $\bigwedge^2
\RR^9\oplus \bigwedge^3 \RR^9 \equiv \so(9)\oplus \bigwedge^3 \RR^9$. In particular, given
a Cartan subalgebra $\mathfrak{h}\subset\so(16)$ (generated by skew-diagonal matrices with
purely imaginary skew eigenvalues), there is a decomposition $\mathfrak{h} =
\mathfrak{h}_2\oplus \mathfrak{h}_3$, where $\mathfrak{h}_2\subset \so(9)$ and
$\mathfrak{h}_3\subset \bigwedge^3 \RR^9$. Since the $\so(16)$ has rank 8 and $\so(9)$ has
rank 4, we can associate $n\leq 4$ 2-form generators and $8-n$ 3-form generators to every
Cartan subalgebra $\mathfrak{h}$ via the isomorphism.

Now obviously
\begin{eqnarray*}
  \left[\mathfrak{h}_2,\mathfrak{h}_2\right] & \subset & \mathfrak{h}_2~,\\
  \left[\mathfrak{h}_2,\mathfrak{h}_3\right] & \subset & \mathfrak{h}_3~,
\end{eqnarray*}
since elements of $\mathfrak{h}_2$ act as infinitesimal $SO(9)$-rotations. Note that this
doesn't necessarily imply that $\mathfrak{h}_3$ is commutative.

But if we further assume that $\mathfrak{h}_3$ is also a Cartan subalgebra (so that
$\left[\mathfrak{h}_3,\mathfrak{h}_3\right] = 0$ and hence
$\left[\mathfrak{h}_,\mathfrak{h}_3\right] = 0$ as well), a direct calculation
\cite{ohta04} shows that only cases that occur are $n=1$ and $n=3$, and a convenient
choices for 2-form and 3-form generators in terms of gamma matrix monomials are
\begin{eqnarray*}
  & & \Gamma_{12},~\Gamma_{34},~\Gamma_{56},~\Gamma_{78}\\
  & & \Gamma_{129},~\Gamma_{349},~\Gamma_{569},~\Gamma_{789}
\end{eqnarray*}
in the $n=1$ case and
\begin{eqnarray*}
  & & \Gamma_{78}\\
  & & \Gamma_{123},~\Gamma_{145},~\Gamma_{167},~\Gamma_{246},~\Gamma_{257},~\Gamma_{347},~\Gamma_{356}
\end{eqnarray*}
in the $n=3$ case.

These two orbit types give rise to 3-form ans\"{a}tze whose action on (complexified)
spinors can be diagonalised. The ansatz for $\Theta$ is then
\begin{equation}
  \label{eq:fourparam}
  \Theta = \alpha_1 dx^{129} + \alpha_2 dx^{349} + \alpha_3 dx^{569} +
  \alpha_4 dx^{789}
\end{equation} or
\begin{equation}
  \label{eq:7param}
  \Theta = \beta_1 dx^{123} + \beta_2 dx^{145} +
  \beta_3 dx^{167} + \beta_4 dx^{246} + \beta_5 dx^{257} +
  \beta_6 dx^{347} + \beta_7 dx^{356}~,
\end{equation} 
where the $\alpha$'s and $\beta$'s are real parameters. As pointed out in
\cite{cvetic_lu_pope02}, if the $\alpha$'s are set to be equal, $\Theta$ is proportional
to $dx^{9}\wedge\omega$, where $\omega$ is the K\"{a}hler form on $\RR^{8}$. Similarly, if
the $\beta$'s agree in the second ansatz, $\Theta$ is proportional to the
$G_{2}$-invariant associative 3-form on $\RR^{7}$. In both cases each of the three-form
terms $\Gamma_{i_1 i_2 i_3}$ for $i_1,i_2,i_3\in \{1\ldots 9\}$ is a complex structure on
the spinor bundle, so when diagonalised they act as $\pm i$. The skew eigenvalues of
$\Theta$ in the four-parameter case are $i\lambda_{a},a = 1\ldots 8$, where
\begin{eqnarray*}
  \lambda_1 & = &  \alpha_1 - \alpha_2 + \alpha_3 - \alpha_4 \\
  \lambda_2 & = & \alpha_1 + \alpha_2 - \alpha_3 - \alpha_4 \\
  \lambda_3 & = & \alpha_1 + \alpha_2 + \alpha_3 - \alpha_4 \\
  \lambda_4 & = & -\alpha_1 -\alpha_2  - \alpha_3 - \alpha_4 \\
  \lambda_5 & = & - \alpha_1 + \alpha_2 + \alpha_3 + \alpha_4 \\
  \lambda_6 & = & \alpha_1 - \alpha_2 - \alpha_3 + \alpha_4 \\
  \lambda_7 & = & \alpha_1 - \alpha_2 + \alpha_3 + \alpha_4 \\
  \lambda_8 & = & - \alpha_1 + \alpha_2 - \alpha_3 + \alpha_4
\end{eqnarray*}
Similarly the skew eigenvalues in the seven-parameter case are given by
$i\lambda^{\prime}_a,a=1\ldots 8$, where
\begin{eqnarray*}
\lambda_{1}^{\prime} & = & -\beta_1  -\beta_2  -\beta_3  -\beta_4 + \beta_5 + \beta_6 + \beta_7 \\
\lambda_{2}^{\prime} & = &  - \beta_1 + \beta_2 + \beta_3 + \beta_4  -\beta_5 + \beta_6 + \beta_7 \\
\lambda_{3}^{\prime} & = &  \beta_1 + \beta_2 - \beta_3 - \beta_4 - \beta_5 -\beta_6 + \beta_7 \\
\lambda_{4}^{\prime} & = &  \beta_1 - \beta_2 + \beta_3 + \beta_4 + \beta_5  -\beta_6 + \beta_7 \\
\lambda_{5}^{\prime} & = &  \beta_1 - \beta_2 + \beta_3 - \beta_4 - \beta_5 + \beta_6 - \beta_7 \\
\lambda_{6}^{\prime} & = &  \beta_1 + \beta_2 - \beta_3 + \beta_4 + \beta_5 + \beta_6 -\beta_7\\
\lambda_{7}^{\prime} & = &  \beta_1 + \beta_2 + \beta_3 - \beta_4 + \beta_5- \beta_6 - \beta_7 \\
\lambda_{8}^{\prime} & = &  -\beta_1 - \beta_2 - \beta_3 + \beta_4 - \beta_5 - \beta_6 - \beta_7
\end{eqnarray*}
Note that by choosing suitable $\alpha$'s or $\beta$'s, some of the eigenvalues can be
made to vanish,
i.e. $\Theta$ can have a nontrivial kernel. Equation \eqref{eq:KSeqn} then implies that
the subspace of Killing spinors that lies in $\Ker\Theta$ will be independent of $x^{-}$. 

We can now work out the possible $A$ that can occur in these ans\"{a}tze, using the
procedure explained in \cite{cvetic_lu_pope02}. In the
4-parameter case, $X_{9} = 4\Theta$ acting on $\chi_{-}$, and thus
\begin{equation}
  \label{eq:mu9}
  \mu^2_{9} =
  \frac{1}{9}\lambda_a^2
\end{equation}
for some choice of $\lambda_{a}$. That is, \emph{a priori} we can
choose $\chi_{-}$ to be any eigenspinor of $\Theta$, and this choice in turn determines
the rest of the $\mu_i$. For example, consider the direction $i=1$. To
determine $\mu_1$, we need to solve the equation
$\left(X_1-\kappa_1\right)\Gamma_{1}\chi_{-}$. Substituting $X_1 = \Gamma_1\Theta\Gamma_1
+ 3\Theta$, we find that
\begin{equation}
  \label{eq:x1}
  \lambda_{a}\chi_{-} + 3\Gamma_{1}\Theta\Gamma_{1}\chi_{-} -
  \kappa_1\chi_{-} = 0~.
\end{equation}
Looking at the form of $\Theta$, it is easy to see that the
eigenvalues of $\Gamma_1\Theta\Gamma_1$ obtained from those of $\Theta$ by reversing the
signs of $\alpha_2,~\alpha_3$ and $\alpha_4$ (since $\Gamma_1$ anticommutes with these
terms). Applying this to $\lambda_a$ is sufficient to solve the previous equation. 
Following this procedure we can solve the rest of the $\mu_i$. A similar argument works in
the 7-parameter case: now $X_8 = X_9 = 2\Theta$. The possible metrics that can occur can
be found in Appendix \ref{sec:metrics}. Note that in the four-parameter case $\mu_1^2 =
\mu_2^2,~\mu_3^2 = \mu_4^2,~\mu_5^2 = \mu_6^2,~\mu_7^2 = \mu_8^2$ and in the
four-parameter case $\mu_8^2 = \mu_9^2$.

The degeneracy of eigenspinors of $\Theta$ satisfying \eqref{eq:chiminus} gives the dimension of
supernumerary Killing spinors. In the generic case where the coefficients of $\Theta$ are
arbitrary there are 2 supernumerary Killing spinors, but there can be more if the
coefficients are chosen so that some of the $\lambda_a$'s or $\lambda^{\prime}_a$'s
agree. The conditions for degeneracy are worked out in detail in \cite{ChrisJerome}. 

In both cases the supernumerary Killing spinors are independent of $x^{-}$ if and only if
$\mu_{9} = 0$. Furthermore, if one or more of the $\mu_{i}$ vanish, the Killing spinors
will be independent of the corresponding transverse coordinates $x^{i}$. For these
solutions the metric will be decomposable: the product of a lower-dimensional
pp-wave with an Euclidean space.

\section{Symmetric discrete quotients of Hpp-waves} \label{sec:isom}
	
Considering all possible quotients of Hpp-solutions by discrete subgroups of
$I(M,g,F)\subset I(M,g)$ (the subgroup of the isometry group of $M,g$ that also preserves
the four-form $F$) is somewhat intractable since we have no classification of the
crystallographic subgroups of Hpp-wave isometry groups at hand. Therefore, we will
restrict ourselves to quotients that are also symmetric. It is known \cite{neukirchner03}
that a quotient of a symmetric space $M = G/H$ (where $G$ is a Lie group and $H$ is the
stabiliser subgroup of a point) by a discrete subgroup $D\subset I(M,g)$ is also symmetric
if and only if $D$ lies in the centraliser of $I(M,g)$ inside the transvection group $G$.
In other words, we want to study quotients by discrete subgroups $D\subset Z$, where
\begin{equation}
  \label{eq:centraliser}
  Z = \{ x\in I(M,g) \mid xh = hx ~\forall h\in G\}~.
\end{equation}
The isometries and conformal symmetries of Cahen-Wallach spaces were investigated by Cahen
and Kerbrat in \cite{cahen_kerbrat}. They also give expressions for the centralisers that
can occur.

There are two possibilities: 
\begin{case}
One of the eigenvalues $\mu_{i}>0$ for some $i$, \emph{or}
$\frac{\mu_{i}}{\mu_{j}}\not\in\mathds{Q}$ for some $(i,j)$. Then
\begin{equation}
  \label{eq:1stcentre} Z\simeq \RR = \{\gamma_{\alpha} \mid
  \gamma_{\alpha}\left(x^{+},x^{-},x^{i}\right) =
  \left(x^{+}+\alpha,x^{-},x^{i}\right)~,
\end{equation} where $\alpha\in\RR$.
\end{case}

\begin{case}
All the eigenvalues $\mu_{i}$ are negative and
$\frac{\mu_{i}}{\mu_{j}}\in \mathds{Q}$ for all $(i,j)$. We write $\mu_{i} =
-k_{i}^{2}$ for all $i$. Then
\begin{equation}
  \label{eq:2ndcentre} Z = \{\gamma_{\alpha,\underline{m}} \mid
  \gamma_{\alpha,\underline{m}}\left(x^{+},x^{-},x^{i}\right) =
  \left(x^{+}+\alpha, x^{-}+\beta, (-1)^{m_i}x^i\right)\}~,
\end{equation} where $m_{i}\in\mathds{Z}$ and $\beta = \frac{\pi m_{i}}{k_{i}}$
for all $i$. 
\end{case}

The ratio of $m_{i}$ and $k_{i}$ is the same for all $i$, and for any $(i,j)$ we can write
\begin{equation}
  \label{eq:mrelation}
  m_{i} = \frac{k_{i}m_{j}}{k_{j}}~.
\end{equation}
The values of all $m_{i}$ are determined by any one of them, so in fact $Z\simeq
\mathds{Z}\oplus \RR$ in this case. Also note that for $N$ to be orientable,
$\sum_{i=1}^{9}m_{i}$ must be even: otherwise the volume form $\dvol = dx^+\wedge dx^-
\wedge dx^1 \wedge\dots\wedge dx^9$ would not be left invariant.

Qualitatively speaking, in all cases quotienting by the action of $Z$ consists of periodic
identifications of the light-cone coordinates and $\mathds{Z}_{2}$-orbifoldings of the
transverse coordinates. We observe that all the pp-wave solutions with supernumerary
supersymmetries are examples of Case 2, provided that the ratios of the coefficients
appearing in $\Theta$ are also rational. 

We are also interested in spinors (in particular Killing spinors) that are left invariant
by the quotient. In Case 2, a discrete subgroup $D_{\alpha,\underline{m}}\subset Z$ is
generated by the elements $\gamma_{\alpha,0}$ and $\gamma_{0,\underline{m}}$. Their
derivatives acting on the frame bundle are
\begin{eqnarray*}
   d\gamma_{\alpha,0} & = & \Id_{11\times 11}~,\\
d\gamma_{0,k} & = & \left(
    \begin{array} {ccccccc} 1 & 0 & 0 & 0 & 0 & \cdots & 0 \\ 0 & 1 & 0 & 0 & 0
& \cdots & 0 \\ 0 & 0 & (-1)^{m_1} & 0 & 0 &\cdots & 0\\ 0 & 0 & 0 & (-1)^{m_2}
& 0 & \cdots & 0 \\ \vdots & 0 & 0 & 0 & \ddots & \cdots & \vdots \\ 0 & 0 & 0 &
0 & 0 &\cdots & (-1)^{m_{9}} \\
    \end{array} \right)
\end{eqnarray*}
Now suppose that $m_{s_{1}},m_{s_2}\ldots,m_{s_{r}}$ are the odd $m_{i}$. The fact that we
want $D_{\alpha,\underline{m}}$ to preserve orientation means that $r$ is even, as we
mentioned previously. Then the quotient $N = M/D_{\alpha,\underline{m}}$ has four possible
spin structures, corresponding to the two possible lifts of $d\gamma_{0,\underline{m}}$:
$\Gamma_{0,\underline{m}} = \pm \Gamma_{s_{1}s_{2}\ldots s_{r}}\in\Spin(9)$ and
$\Gamma_{\alpha,0} = \pm\Id$ \cite{baum00}. Using these expressions and equation
\eqref{eq:spinorfields}, we can then study the existence of invariant spinors explicitly. 

It is easy to see that quotients of solutions for which Case 1 applies are rather trivial:
there are only two possible spin structures and since generic Killing spinors do not
depend on $x^{+}$, only the trivial lift of $\gamma_{\alpha,0}$ will preserve any (and
in fact all) Killing spinors. Therefore, in the sequel we will focus on the solutions that
admit supernumerary Killing spinors.

\subsection{The maximally supersymmetric case} \label{sec:maxsusy}

To begin with a simple example before studying the generaly supernumerary case in detail,
let us analyse the symmetric discrete quotients of the maximally supersymmetric solution
\eqref{eq:maxsusy} and see which choices of spin structure preserve Killing spinors. Now
$A_{ij}$ is diagonal and all eigenvalues of are negative,so Case $2$ above applies. To
obtain a quotient isometric to $(M,g,F)$ as a supergravity solution, we want to focus on a
subgroup $Z_{F}\subset Z$ that also preserves the four-form $F$. Looking at the form of
$F$ in \eqref{eq:maxsusy}, we observe that an element $\gamma_{\alpha,\underline{m}}\in
Z$ will preserve $F$ if and only if \emph{none} of $(m_1,m_2,m_3)$ are odd or if
\emph{two} of them are: $\gamma_{\alpha,\underline{m}}$ acts trivially on $dx^{-}$. Now
\begin{equation*} k_{i} =
  \begin{cases} \frac{1}{3}\mu ~, i=1\ldots 3\\ \frac{1}{6}\mu~, i=4\ldots 9
  \end{cases}
\end{equation*}
But since the $k_{i}$ are equal for $i=1\ldots 3$, equation \eqref{eq:mrelation} implies
that $m_1 = m_2 = m_3$. Therefore, for $\gamma_{0,\underline{m}}$ to preserve $F$, $m_1 =
m_2 = m_3 \equiv 2k$ for some $k\in\mathds{Z}$. Equation \eqref{eq:mrelation} also implies
that $m_4 = \ldots = m_9 \equiv k$. We conclude that
\begin{equation}
  \label{eq:zF} Z_{F} = \{\gamma_{\alpha,k}\in Z \mid
  \gamma_{\alpha,k}(x^{+},x^{-},x^{i}) =
  \left(x^{+}+\alpha,x^{-}+\beta,x^{1,2,3},(-1)^kx^{4,\dots,9}\right)\}~,
\end{equation}
where $\beta = \frac{6\pi k}{\mu}$ and $\alpha\in\RR$.

A discrete subgroup $D_{\alpha,k}$ of $Z_{F}$ is generated by elements
$\gamma_{\alpha,0}$ and $\gamma_{0,1}$. Looking at the condition
\eqref{eq:spinorfields}, it is again obvious that if $d\gamma_{\alpha,0} $ lifts to
$-\Id$, no spinors will be left invariant. But provided that $\Gamma_{\alpha,0} = \Id$,
there are two possible spin structures depending on the choice of sign for
$\pm\Gamma_{0,1}$.

The derivative of $\gamma_{0,1}$ lifts to the spinor bundle as $\pm\Gamma_{4\ldots 9}$.
Using the familiar trigonometric identities $\cos\left(\tfrac{\pi}{2} -\theta\right) =
-\sin\theta$ and $\sin\left(\tfrac{\pi}{2} -\theta\right) = \cos\theta$, we obtain
\begin{multline}
  \label{eq:psiplus} \varepsilon_{\psi_+,\psi_-}\left(\gamma_{(0,1)}(x)\right)
  = \left(-\cos\left(\frac{\mu}{4}x^{-}\right)\Id+
    \sin\left(\frac{\mu}{4}x^{-}\right)I\right)\psi_{+}\\
  +\left(\cos\left(\frac{\mu}{4}x^{-}\right)I +
    \sin\left(\frac{\mu}{4}x^{-}\right)\Id\right)\psi_{-}\\
  +\tfrac{1}{6}\mu\left(\sum_{i\leq 3}^{}x^{i}\Gamma_{i} + \half\sum_{\geq
      4}x^{i}\Gamma_{i}\right)\left(\cos\left(\frac{\mu}{4}x^{-}\right)\Id +
    \sin\left(\frac{\mu}{4}x^{-}\right)I\right)\Gamma_{+}\psi_{-}~,
\end{multline}
Comparing this expression with \eqref{eq:KSeqn} and noting that $\Gamma_{{0,1}}$
anticommutes with $\Gamma_{i}$ for $i=4\ldots 9$, we find that we can write this as
\begin{equation*}
  \varepsilon_{\psi_+,\psi_-}\left(\gamma_{0,1}(x)\right) = \varepsilon_{-I\psi_+,I\psi_-}(x)~.
\end{equation*}
Thus, the action of $\gamma_{(0,1)}$ leaves $\varepsilon$ invariant if
\begin{eqnarray*}
  \Gamma_{0,1}\psi_{+} = -I\psi_{+}~,\\
  \Gamma_{0,1}\psi_{-} = I\psi_{-}~,
\end{eqnarray*}
Recall that we have chosen the spinor module for which the action of the centre of the
Clifford algebra (and thus the volume element) agrees with $-\Id$. Then
\begin{equation}
  \label{eq:clcenter}
  \Gamma_{-+}\Gamma_{1\ldots 9}\psi_{+} = -\psi_{+}~,
\end{equation}
implies that (since $\Gamma_{-+} = \Gamma_{-}\Gamma_{+} + \Id$)
\begin{equation*}
  \Gamma_{4\ldots 9}\psi_{+} = I\psi_{+}~.
\end{equation*}
Correspondingly, $\Gamma_{4\ldots 9}\psi_{-} = -I\psi_{-}$. Thus, equation
\eqref{eq:psiplus} is satisfied if and only if $\Gamma_{0,1} = -\Gamma_{4\ldots 9}$. We
can therefore conclude that for all symmetric discrete quotients of the maximally
supersymmetric Hpp-wave, out of the four possible choices of spin structure there is
precisely one that preserves any (and indeed, all) supersymmetry.

\subsection{The four-parameter case}
\label{sec:4param}

Going through all possible quotients of supernumerary pp-wave solutions listed in Appendix
\ref{sec:metrics} using the explicit form of the Killing spinors would be somewhat
tedious, so instead we use a method that can be implemented more easily using a computer
program (in our case, a Mathematica notebook).

To study the moduli space of possible quotients, we could take the coefficients of
$\Theta$ as the data, allow them to vary and examine the consequences, as in
\cite{ChrisJerome}. However, for computational purposes it is actually more useful to take
the 9-tuple $(m_1,\ldots,m_9)$ and the eigenvalue $\lambda_{a_0}$ (where $\lambda_{a_0}$
is the eigenvalue chosen to appear in equation \eqref{eq:mu9}, that is, $\mu_{9} =
\frac{1}{9}\lambda_{a_{0}}^2$) as the data defining a quotient. Looking at the different
metrics appearing in Table \ref{tab:fourparameter}, we find that it is always possible to
express the $\alpha$'s --- and thus the eigenvalues $\lambda_{a}$ --- in terms of $k_i$
(recall that the $k_i$ are related to the eigenvalues of the matrix $A$ by $\mu_{i} =
-k_i^2$). In other words, we can write $\lambda_{a} = \sum_i^{}c_{i}k_{i}$ for each
$\lambda_a$ and for some coefficients $c_i$. Given $(m_1,\ldots,m_9)$, we can use equation
\eqref{eq:mrelation} to determine the $k_i$ and hence the coefficients
$\alpha_1,\ldots,\alpha_4$. Knowing $m_1\ldots,m_9,\lambda_{a_0}$ for the quotient is thus
sufficient to determine the original solution.

Restricting to $Z_{F}$, the subgroup of $Z$ that also preserves the four-form $F$, we
observe that $m_9$ must always be even. Using the equation \eqref{eq:mrelation} and the
remarks in section \ref{sec:cwm}, we also know that $m_1 = m_2,~m_3 = m_4,~m_5 = m_6$ and
$m_7 = m_8$. To preserve orientation, $\sum_i^{}m_i$ has to be even as well, but in this
case this imposes no further restrictions, since there is always an even number of odd
$m_i$. 

It is convenient to express the Killing spinors using the eigenspinors of $\Theta$ as a
basis. Note that acting on the $\lambda_{a}$-eigenspace, $J_{\lambda_{a}} =
\tfrac{1}{\lambda_{a}}\Theta$ is a complex structure. Thus, we can write the
exponentials appearing in $\chi_{\pm}$ explicitly as
\begin{eqnarray}
  \label{eq:spinorbasis}
  \chi_{+} = \sum_{a=1}^{8} \left(\cos\left(\tfrac{\lambda_{a}x^{-}}{4}\right) +
    i\sin\left(\tfrac{\lambda_{a}x^{-}}{4}\right)\right)\psi_{+}(\lambda_{a})~,\\
\chi_{-} = \left(\cos\left(\tfrac{\lambda_{a_{0}}x^{-}}{12}\right) +
    i\sin\left(\tfrac{\lambda_{a_{0}}x^{-}}{12}\right)\right)\psi_{+}(\lambda_{a_{0}})  
\end{eqnarray}
where $\Theta\cdot\psi_{\pm}(\lambda_{a}) = i\lambda_{a}\psi_{\pm}(\lambda_{a})$.

To determine the fraction of supersymmetry preserved by a symmetric discrete quotient, we
need to work out how $\gamma_{0,\underline{m}}$ and $\Gamma_{0,\underline{m}}$ act on
$\chi_{\pm}$. Recall that $\gamma_{0,\underline{m}}(x^{-}) = x^{-} +
\frac{m_{i}}{k_{i}}\pi$ for some $i$. Computing the action of this shift on $\chi_{\pm}$
is straightforward. As mentioned above, we can express the $\alpha$'s --- and thus the
eigenvalues $\lambda_{a}$ --- in terms of $k_i$. Thus, 
\begin{eqnarray}
  \label{eq:lambdashift}
  \lambda_{a}\frac{m_{j}}{k_{j}} & =  & \sum_{i}^{}c_{i}\frac{m_j k_i}{kj}\\
  & = & \sum_{i}^{}c_{i}m_{i}~, 
\end{eqnarray}
so under the action of the isometry, $\lambda_{a}x^{-}\rightarrow \lambda_a x^{-} +
\pi\sum_{i}^{}c_i m_i$. Using this observation and usual trigonometric identities, we can
work out how the trigonometric functions in \eqref{eq:spinorbasis} transform under
$\gamma_{0,\underline{m}}$. 

We also need to know how $\Gamma_{0,\underline{m}}$ acts on the eigenspinors. Since we're
taking the $m_i$ to be our data, it is not hard to enumerate the possibilities. In the
four-parameter case, each of $m_1,~m_3,~m_4,~m_5$ and $m_7$ can be even or odd.

If we write the 3-form in this ansatz as $\Theta = \alpha_1 I_1 + \alpha_2 I_2 +\alpha_3
I_3 + \alpha_4 I_4$, we can express any $\lambda$ as
\begin{equation}
\label{eq:lambda_action}
  \lambda = \epsilon_1(\lambda)\alpha_1 + \epsilon_2(\lambda)\alpha_2 +\epsilon_3(\lambda)\alpha_3 + \epsilon_4(\lambda)\alpha_4~,
\end{equation}
where $I_p \psi_{\pm}(\lambda) = i\epsilon_{p}(\lambda)\psi_{\pm}(\lambda),~p=1\ldots 4$
and $\epsilon_p(\lambda) = \pm 1$. It is not hard to see that any
$\Gamma_{0,\underline{m}}$ can be written as a product of the $I_p$'s or identified with
such a product via the identity $\Gamma_{1\ldots 9}\psi_{\pm} = \pm\psi$ that relates the
Clifford action of a form on $\RR^9$ to that of its dual. We can thus always write
\begin{equation*}
  \Gamma_{0,\underline{m}} = \epsilon(q)I_{p_1}I_{p_2}\cdot\ldots\cdot I_{p_{q}}
\end{equation*}
acting on $\psi_{+}$ and 
\begin{equation*}
    \Gamma_{0,\underline{m}} = -\epsilon(q)I_{p_1}I_{p_2}\cdot\ldots\cdot I_{p_{q}}~
\end{equation*}
acting on $\psi_{-}$, where $1\leq q \leq 4$ and $\epsilon(q) = -1$ if $q=1$ and $1$
otherwise.

Since the action of each of the $I_p$ on $\psi_{\pm}(\lambda)$ is fixed by equation
\eqref{eq:lambda_action}, the action of $\Gamma_{0,\underline{m}}$ on
$\psi_{\pm}(\lambda)$ is given by
\begin{equation*}
  \Gamma_{0,\underline{m}}\cdot\psi_{\pm}(\lambda) = \mp\epsilon(q)\epsilon_{p_1}\epsilon_{p_2}\ldots\epsilon_{p_q}i^q\psi_{+}(\lambda)~,
\end{equation*}
With these observations, the problem becomes essentially algorithmic and can be easily
implemented in a symbolic computation environment. We have written a Mathematica
notebook\footnote{Available upon request from the author.} that goes through the elements
$\gamma_{\alpha,\underline{m}}$ that generate discrete subgroups $D$ and computes their
action on the Killing spinors. It turns out that for every quotient, the result is similar
to what occurs in the maximally supersymmetric case: out of the four possible spin
structures, there is only one that preserves any of the original Killing spinors.

\subsection{The seven-parameter case}
\label{sec:7param}

The method we described in the previous section also works in the seven-parameter case,
but now we must take care to ensure that the discrete subgroups $D$ also preserve the
four-form $F$. The most convenient way to express this condition is to require that for
each term $\Gamma_{ijk}$ appearing in $\Theta$, $m_i+m_j+m_k$ must be even. In other
words, we have the equations
\begin{eqnarray*}
  \label{eq:zfeqns}
  m_1 + m_2 + m_3 & = &  0\\
  m_1 + m_4 + m_5 & = &  0\\
  m_1 + m_6 + m_7 & = &  0\\
  m_2 + m_4 + m_6 & = &  0\\
  m_2 + m_5 + m_7 & = &  0\\
  m_3 + m_4 + m_7 & = &  0\\
  m_3 + m_5 + m_6 & = & 0~.
\end{eqnarray*}
modulo $2$. This system of equations is not hard to solve over $\ZZ_2$, and thus we find
that the possible forms that $\Gamma_{0,\underline{m}}$ can take are:
\begin{eqnarray*}
& &   \pm\Gamma_{1 2 4 7},~\pm\Gamma_{1 2 5 6},~\pm\Gamma_{1 3 4 6}\\
& &   \pm\Gamma_{1 3 5 7},~\pm\Gamma_{2 3 4 5},~\pm\Gamma_{2 3 6 7},~\pm\Gamma_{4 5 6 7}
\end{eqnarray*}
Again, we note that all these terms can be written as products of terms in $\Theta$, and
thus the method described in the previous section works, provided that we take the
above constraints into account.

Examining all the possible quotients yields the same result as in the four-parameter case:
for all possible quotients, there is only \emph{one} choice of spin structure that preserves any
(and indeed all) of the supersymmetry of the original background. Furthermore, we do not
obtain any new fractions of supersymmetry in either case.

\section{A conjecture}
\label{sec:conjecture}

The results of the previous section lead to the curious observation that all symmetric
quotients of known symmetric M-theory backgrounds with more than 16 Killing spinors
possess a unique spin structure that preserves supersymmetry -- in contrast to the
supersymmetric space forms described in \cite{figueroa-ofarrill:_super_and_spin_struc}.
The only such backgrounds we haven't yet considered are Freund-Rubin -type solutions of
the form $\AdS\times S/\ZZ_2$, since the only symmetric spherical space form is the
projective space\cite{wolf77}, and it is easy to see that there is no ambiguity about the
choice of spin structure in this case -- this situation only arises if $|D|\geq 4$,
where $D$ is the discrete group used in the
quotient\cite{figueroa-ofarrill:_super_and_spin_struc}. \FIXME{Details about this?} Thus
we arrive at a

\begin{conj}
  All symmetric quotients of symmetric M-theory backgrounds for which $\nu>\half$ possess
  a unique spin structure which preserves all of the original supersymmetry.  
\end{conj}

We now show that the requirement $\nu>\half$ is in fact necessary. 

Provided that we drop the requirement of supernumerary Killing spinors, it is not hard to
exhibit examples of symmetric quotients of Hpp-waves that admit more than one spin
structure preserving some Killing spinors. For example, consider a solution of the form
\begin{eqnarray}
  \label{eq:24soln} \Theta & = & \mu\left(dx^{129} + dx^{349}\right)\\ A_{ij} &
= &
  \begin{cases} -\frac{\mu^2}{9}\delta_{ij},~ i,j = 1,\ldots 4\\
-\frac{\mu^2}{36}\delta_{ij},~ i,j = 5\ldots 8\\ -\frac{4\mu^2}{9},~ i=j=9~.
  \end{cases}
\end{eqnarray}
The solution is obtained by taking a solution preserving 24 supersymmetries given by
taking $\alpha_1 = \alpha_2,~\alpha_3 = \alpha_4 = 0$ in the four-parameter case and
permuting the values of the $k_i$: equation \eqref{eq:chiminus} is no longer satisfied,
and thus this solution only admits 16 supersymmetries.

Let us analyse the centraliser of the isometry group. Now $k_1 = k_2 = k_3 = k_4 =
\frac{\mu}{3}$, $k_5 = \ldots = k_8 = \frac{\mu}{6}$ and $k_9 = \frac{2\mu}{3}$. Equation
\eqref{eq:mrelation} then implies that $m_1 = \ldots = m_4$ and $m_5 = \ldots = m_8\equiv
m$. Moreover, $m_1 = 2m_{5}$ and $m_9 = 4m_5$: in other words, $m_1,\ldots m_4$ and $m_9$
will always be even, and since they correspond to the transverse directions that appear in
the form of $\Theta$, all elements of $Z$ will preserve $F$ as well. Thus, $Z$ is of the
form
\begin{equation}
  \label{eq:supernum}
  Z = \{\gamma_{\alpha,k}\in Z \mid\gamma_{\alpha,k}(x^{+},x^{-},x^{i}) =
  \left(x^{+}+\alpha,x^{-}+\beta,x^{1,2,3,4},(-1)^kx^{5,\dots,8},x^9\right)\}~,
\end{equation}
Again, $Z$ is generated by $\gamma_{\alpha,0}$ and $\gamma_{0,1}$. These elements lift to
the spinor bundle as $\Gamma_{\alpha,0} = \pm\Id$ and $\Gamma_{0,1} = \pm\Gamma_{5678}$.
The Killing spinors are of the form given in equation \eqref{eq:cwks}. We could, of
course, use the method described in the previous section and decompose the spinorial
parameter $\psi_{+}$ into eigenspinors of $\Theta$ and work out precisely how
$\Gamma_{0,1}$ acts on them, but it is sufficient to observe that
\eqref{eq:spinorfields} becomes
\begin{equation*}
  \varepsilon_{\psi_{+}}\left(\gamma_{0,1}(x)\right) =
\varepsilon_{\Gamma_{1234}\psi_{+}}(x)~.
\end{equation*}
For this equation to be satisfied we must have $\Gamma_{5678}\psi_{+} =
\pm\Gamma_{1234}\psi_{+}$. In other words, $\psi_{+}$ must lie in the $\pm$-eigenspaces of
$\Gamma_{12\ldots 8}$, depending on which lift of $\gamma_{0,1}$ we choose. Half of the
$\psi_{+}$ satisfy this additional condition; $\Gamma_{12\ldots 8}$ commutes with
$X^{2}_{i}$, so demanding that its eigenspinors belong to the $\pm$-eigenspace of
$\Gamma_{12\ldots 8}$ is an independent constraint.

We conclude that out of the four possible spin structures on the quotient, two admit no
Killing spinors and two preserve 8 of the original sixteen supersymmetries, thus showing
that the inequality in our conjecture must be sharp.

\section{Conclusions}
\label{sec:concs}

In this paper we have examined the symmetric discrete quotients of all the known symmetric
M-theory backgrounds with more than 16 Killing spinors. In all cases, we found that there
is a unique spin structure that preserves all of the original supersymmetry. It would be
interesting to study the interplay between spin structure, symmetry and supersymmetry and
find a formal proof of the conjecture presented in section \ref{sec:conjecture}. Failing
that, as we mentioned in section \ref{sec:cwm}, the known Hpp-wave solutions with
supernumerary supersymmetries are very special and a more careful study of the moduli
space of these solutions might reveal loci for which the matrix $A$ is not diagonal but
which still admit more than 16 Killing spinors. It would be interesting to see if these
solutions could provide counterexamples to our conjecture.

\bibliographystyle{utphys} \bibliography{spin}

\appendix
\newpage
\section{Metrics of pp-waves with supernumerary supersymmetries}
\label{sec:metrics}
\newpage
\scriptsize{
  \begin{table}
\label{tab:fourparameter}
    \centerline{
      \begin{tabular} {|c|l|} \hline $\mu_{9}$ & $\mu_{i}$,$i=1\ldots 8$\\
        \hline \hline $-\frac{1}{9}\left(\alpha_1 + \alpha_2 +\alpha_3 +
          \alpha_4\right)^2$ & $\mu_1 = \mu_2 =
        -\frac{1}{36}(\alpha_1-\alpha_2-\alpha_3-\alpha_4)^2$ \\ & $\mu_3 = \mu_4 =
        -\frac{1}{36}(\alpha_1-2\alpha_2+\alpha_3+\alpha_4)^2$ \\ & $\mu_5 = \mu_6 =
        -\frac{1}{36}(\alpha_1+\alpha_2-2\alpha_3+\alpha_4)^2$ \\ & $\mu_7 = \mu_8 =
        -\frac{1}{36}(\alpha_1+\alpha_2+\alpha_3-2 \alpha_4)^2$ \\ \hline
        $-\frac{1}{9}(-\alpha_1 -\alpha_2 + \alpha_3 -\alpha_4)^2$ & $\mu_1 = \mu_2 =
        -\frac{1}{36}(2 \alpha_1-\alpha_2+\alpha_3-\alpha_4)^2$ \\ & $\mu_3 = \mu_4 =
        -\frac{1}{36}(\alpha_1 -2\alpha_2-\alpha_3+\alpha_4)^2$ \\ & $\mu_5 = \mu_6 =
        -\frac{1}{36} (\alpha_1+\alpha_2+2\alpha_3+\alpha_4)^2$ \\ & $\mu_7 = \mu_8 =
        -\frac{1}{36} (\alpha_1+\alpha_2-\alpha_3-2 \alpha_4)^2$ \\ \hline
        $-\frac{1}{9}(\alpha_1 +\alpha_2 -\alpha_3 -\alpha_4)^2$ & $\mu_1 = \mu_2 =
        -\frac{1}{36}(2 \alpha_1-\alpha_2+\alpha_3+\alpha_4)^2$ \\ & $\mu_3 = \mu_4 =
        -\frac{1}{36}(\alpha_1-2\alpha_2-\alpha_3-\alpha_4)^2$ \\ & $\mu_5 = \mu_6 =
        -\frac{1}{36} (\alpha_1+\alpha_2+2\alpha_3-\alpha_4)^2$ \\ & $\mu_7 = \mu_8 =
        -\frac{1}{36} (\alpha_1+\alpha_2-\alpha_3+2\alpha_4)^2$ \\ \hline
        $-\frac{1}{9}(\alpha_1 + \alpha_2 + \alpha_3 - \alpha_4)^2$ & $\mu_1 = \mu_2
        = -\frac{1}{36}(2\alpha_1-\alpha_2-\alpha_3+\alpha_4)^2$ \\ & $\mu_3 = \mu_4
        = -\frac{1}{36}(\alpha_1-2\alpha_2+\alpha_3-\alpha_4)^2$ \\ & $\mu_5 = \mu_6
        = -\frac{1}{36}(\alpha_1+\alpha_2-2\alpha_3-\alpha_4)^2$ \\ & $\mu_7 = \mu_8
        = -\frac{1}{36}(\alpha_1+\alpha_2+\alpha_3+2 \alpha_4)^2$ \\ \hline
        $-\frac{1}{9}(-\alpha_1 +\alpha_2 + \alpha_3 +\alpha_4)^2$ & $\mu_1 = \mu_2 =
        -\frac{1}{36}(\alpha_1+\alpha_2+\alpha_3+\alpha_4)^2$ \\ & $\mu_3 = \mu_4 =
        -\frac{1}{36}(\alpha_1+2\alpha_2-\alpha_3-\alpha_4)^2$ \\ & $\mu_5 = \mu_6 =
        -\frac{1}{36}(\alpha_1-\alpha_2+2\alpha_3-\alpha_4)^2$ \\ & $\mu_7 = \mu_8 =
        -\frac{1}{36}(\alpha_1-\alpha_2-\alpha_3+2 \alpha_4)^2$ \\ \hline
        $-\frac{1}{9}(\alpha_1 -\alpha_2 - \alpha_3 +\alpha_4)^2$ & $\mu_1 = \mu_2 =
        -\frac{1}{36}(2\alpha_1+\alpha_2+\alpha_3-\alpha_4)^2$ \\ & $\mu_3 = \mu_4 =
        -\frac{1}{36}(\alpha_1+2\alpha_2-\alpha_3+\alpha_4)^2$ \\ & $\mu_5 = \mu_6 =
        -\frac{1}{36}(\alpha_1-\alpha_2+2\alpha_3+\alpha_4)^2$ \\ & $\mu_7 = \mu_8 =
        -\frac{1}{36}(\alpha_1-\alpha_2-\alpha_3-2 \alpha_4)^2$ \\ \hline
        $-\frac{1}{9}(\alpha_1 -\alpha_2 +\alpha_3 + \alpha_4)^2$ & $\mu_1 = \mu_2 =
        -\frac{1}{36}(2\alpha_1+\alpha_2-\alpha_3-\alpha_4)^2$ \\ & $\mu_3 = \mu_4 =
        -\frac{1}{36}(-\alpha_1-2\alpha_2-\alpha_3-\alpha_4)^2$ \\ & $\mu_5 = \mu_6 =
        -\frac{1}{36}(-\alpha_1+\alpha_2+2\alpha_3-\alpha_4)^2$ \\ & $\mu_7 = \mu_8 =
        -\frac{1}{36}(-\alpha_1+\alpha_2-\alpha_3+2\alpha_4)^2$ \\ \hline
        $-\frac{1}{9}(-\alpha_1 +\alpha_2 -\alpha_3 +\alpha_4)^2$ & $\mu_1 = \mu_2 =
        -\frac{1}{36}(-2\alpha_1-\alpha_2+\alpha_3-\alpha_4)^2$ \\ & $\mu_3 = \mu_4 =
        -\frac{1}{36}(\alpha_1+2\alpha_2+\alpha_3-\alpha_4)^2$ \\ & $\mu_5 = \mu_6 =
        -\frac{1}{36}(\alpha_1-\alpha_2-2\alpha_3-\alpha_4)^2$ \\ & $\mu_7 = \mu_8 =
        -\frac{1}{36}(\alpha_1-\alpha_2+\alpha_3+2 \alpha_4)^2$ \\ \hline
      \end{tabular}}
      \caption{Metrics associated to different eigenvalues of
        $\Theta$ for the 4-parameter ansatz}
  \end{table} 

\tiny{
  \begin{table}\label{tab:sevenparam} \centerline{
      \begin{tabular} {|c|l|} \hline $\mu_8 = \mu_9$ & $\mu_i$,~$i=1\ldots
        7$ \\ \hline \hline $-\frac{1}{36}(-\beta_1 -\beta_2 -\beta_3 -\beta_4 +\beta_5
        +\beta_6 +\beta_7)^2$ & $\mu_1 = -\frac{1}{36}(-2 \beta_1-2
        \beta_2-2\beta_3+\beta_4-\beta_5-\beta_6-\beta_7)^2$ \\ & $\mu_2 =
        -\frac{1}{36}(-2 \beta_1+\beta_2+\beta_3-2 \beta_4+2 \beta_5-\beta_6-\beta_7)^2$
        \\ & $\mu_3 = -\frac{1}{36}(-2 \beta_1+\beta_2+\beta_3+\beta_4-\beta_5+2
        \beta_6+2 \beta_7)^2$ \\ & $\mu_4 = -\frac{1}{36}(\beta_1-2 \beta_2+\beta_3-2
        \beta_4-\beta_5+2 \beta_6-\beta_7)^2$ \\ & $\mu_5 = -\frac{1}{36}(\beta_1-2
        \beta_2+\beta_3+\beta_4+2 \beta_5-\beta_6+2 \beta_7)^2$ \\ & $\mu_6 =
        -\frac{1}{36}(\beta_1+\beta_2-2 \beta_3-2 \beta_4-\beta_5-\beta_6+2 \beta_7)^2$
        \\ & $\mu_7 = -\frac{1}{36}(\beta_1+\beta_2-2 \beta_3+\beta_4+2 \beta_5+2
        \beta_6-\beta_7)^2$ \\ \hline $-\frac{1}{36}(-\beta_1 +\beta_2 +\beta_3 +\beta_4
        -\beta_5 +\beta_6 +\beta_7)^2$ & $\mu_1 = -\frac{1}{36}(-2 \beta_1+2 \beta_2+2
        \beta_3-\beta_4+\beta_5-\beta_6-\beta_7)^2$ \\ & $\mu_2 = -\frac{1}{36}(-2
        \beta_1-\beta_2-\beta_3+2 \beta_4-2 \beta_5-\beta_6-\beta_7)^2$ \\ & $\mu_3 =
        -\frac{1}{36}(-2 \beta_1-\beta_2-\beta_3-\beta_4+\beta_5+2 \beta_6+2 \beta_7)^2$
        \\ & $\mu_4 = -\frac{1}{36}(\beta_1+2 \beta_2-\beta_3+2 \beta_4+\beta_5+2
        \beta_6-\beta_7)^2$ \\ & $\mu_5 = -\frac{1}{36}(\beta_1+2
        \beta_2-\beta_3-\beta_4-2 \beta_5-\beta_6+2 \beta_7)^2$ \\ & $\mu_6 =
        -\frac{1}{36}(\beta_1-\beta_2+2 \beta_3+2 \beta_4+\beta_5-\beta_6+2 \beta_7)^2$
        \\ & $\mu_7 = -\frac{1}{36}(\beta_1-\beta_2+2 \beta_3-\beta_4-2 \beta_5+2
        \beta_6-\beta_7)^2$ \\ \hline $-\frac{1}{36}(\beta_1 +
        \beta_2-\beta_3-\beta_4-\beta_5-\beta_6+\beta_7)^2$ & $\mu_1 = -\frac{1}{36}(2
        \beta_1+2 \beta_2-2 \beta_3+\beta_4+\beta_5+\beta_6-\beta_7)^2$ \\ & $\mu_2 =
        -\frac{1}{36}(2 \beta_1-\beta_2+\beta_3-2 \beta_4-2 \beta_5+\beta_6-\beta_7)^2$
        \\ & $\mu_3 = -\frac{1}{36}(2 \beta_1-\beta_2+\beta_3+\beta_4+\beta_5-2
        \beta_6+2 \beta_7)^2$ \\ & $\mu_4 = -\frac{1}{36}(-\beta_1+2 \beta_2+\beta_3-2
        \beta_4+\beta_5-2 \beta_6-\beta_7)^2$ \\ & $\mu_5 = -\frac{1}{36}(-\beta_1+2
        \beta_2+\beta_3+\beta_4-2 \beta_5+\beta_6+2 \beta_7)^2$ \\ & $\mu_6 =
        -\frac{1}{36}(-\beta_1-\beta_2-2 \beta_3-2 \beta_4+\beta_5+\beta_6+2 \beta_7)^2$
        \\ & $\mu_7 = -\frac{1}{36}(-\beta_1-\beta_2-2 \beta_3+\beta_4-2 \beta_5-2
        \beta_6-\beta_7)^2$ \\ \hline $-\frac{1}{36}(\beta_1 -\beta_2 +\beta_3 +
        \beta_4+\beta_5 -\beta_6+\beta_7)^2$ & $\mu_1 = -\frac{1}{36}(2 \beta_1-2
        \beta_2+2 \beta_3-\beta_4-\beta_5+\beta_6-\beta_7)^2$ \\ & $\mu_2 =
        -\frac{1}{36}(2 \beta_1+\beta_2-\beta_3+2 \beta_4+2 \beta_5+\beta_6-\beta_7)^2$
        \\ & $\mu_3 = -\frac{1}{36}(2 \beta_1+\beta_2-\beta_3-\beta_4-\beta_5-2
        \beta_6+2 \beta_7)^2$ \\ & $\mu_4 = -\frac{1}{36}(-\beta_1-2 \beta_2-\beta_3+2
        \beta_4-\beta_5-2 \beta_6-\beta_7)^2$ \\ & $\mu_5 = -\frac{1}{36}(-\beta_1-2
        \beta_2-\beta_3-\beta_4+2 \beta_5+\beta_6+2 \beta_7)^2$ \\ & $\mu_6 =
        -\frac{1}{36}(-\beta_1+\beta_2+2 \beta_3+2 \beta_4-\beta_5+\beta_6+2 \beta_7)^2$
        \\ & $\mu_7 = -\frac{1}{36}(-\beta_1+\beta_2+2 \beta_3-\beta_4+2 \beta_5-2
        \beta_6-\beta_7)^2$ \\ \hline $-\frac{1}{36}(\beta_1 -\beta_2+ \beta_3 -\beta_4
        -\beta_5 + \beta_6 -\beta_7)^2$ & $\mu_1 = -\frac{1}{36}(2 \beta_1-2 \beta_2+2
        \beta_3+\beta_4+\beta_5-\beta_6+\beta_7)^2$ \\ & $\mu_2 = -\frac{1}{36}(2
        \beta_1+\beta_2-\beta_3-2 \beta_4-2 \beta_5-\beta_6+\beta_7)^2$ \\ & $\mu_3 =
        -\frac{1}{36}(2 \beta_1+\beta_2-\beta_3+\beta_4+\beta_5+2 \beta_6-2 \beta_7)^2$
        \\ & $\mu_4 = -\frac{1}{36}(-\beta_1-2 \beta_2-\beta_3-2 \beta_4+\beta_5+2
        \beta_6+\beta_7)^2$ \\ & $\mu_5 = -\frac{1}{36}(-\beta_1-2
        \beta_2-\beta_3+\beta_4-2 \beta_5-\beta_6-2 \beta_7)^2$ \\ & $\mu_6 =
        -\frac{1}{36}(-\beta_1+\beta_2+2 \beta_3-2 \beta_4+\beta_5-\beta_6-2 \beta_7)^2$
        \\ & $\mu_7 = -\frac{1}{36}(-\beta_1+\beta_2+2 \beta_3+\beta_4-2 \beta_5+2
        \beta_6+\beta_7)^2$ \\ \hline $-\frac{1}{36}(\beta_1 + \beta_2 - \beta_3 +\beta_4
        +\beta_5 + \beta_6 -\beta_7)^2$ & $\mu_1 = -\frac{1}{36}(2 \beta_1+2 \beta_2-2
        \beta_3-\beta_4-\beta_5-\beta_6+\beta_7)^2$ \\ & $\mu_2 = -\frac{1}{36}(2
        \beta_1-\beta_2+\beta_3+2 \beta_4+2 \beta_5-\beta_6+\beta_7)^2$ \\ & $\mu_3 =
        -\frac{1}{36}(2 \beta_1-\beta_2+\beta_3-\beta_4-\beta_5+2 \beta_6-2 \beta_7)^2$
        \\ & $\mu_4 = -\frac{1}{36}(-\beta_1+2 \beta_2+\beta_3+2 \beta_4-\beta_5+2
        \beta_6+\beta_7)^2$ \\ & $\mu_5 = -\frac{1}{36}(-\beta_1+2
        \beta_2+\beta_3-\beta_4+2 \beta_5-\beta_6-2 \beta_7)^2$ \\ & $\mu_6 =
        -\frac{1}{36}(-\beta_1-\beta_2-2 \beta_3+2 \beta_4-\beta_5-\beta_6-2 \beta_7)^2$
        \\ & $\mu_7 = -\frac{1}{36}(-\beta_1-\beta_2-2 \beta_3-\beta_4+2 \beta_5+2
        \beta_6+\beta_7)^2$ \\ \hline $-\frac{1}{36}(-\beta_1 + \beta_2 + \beta_3
        -\beta_4 + \beta_5 -\beta_6 -\beta_7)^2$ & $\mu_1 = -\frac{1}{36}(-2 \beta_1+2
        \beta_2+2 \beta_3+\beta_4-\beta_5+\beta_6+\beta_7)^2$ \\ & $\mu_2 =
        -\frac{1}{36}(-2 \beta_1-\beta_2-\beta_3-2 \beta_4+2 \beta_5+\beta_6+\beta_7)^2$
        \\ & $\mu_3 = -\frac{1}{36}(-2 \beta_1-\beta_2-\beta_3+\beta_4-\beta_5-2
        \beta_6-2 \beta_7)^2$ \\ & $\mu_4 = -\frac{1}{36}(\beta_1+2 \beta_2-\beta_3-2
        \beta_4-\beta_5-2 \beta_6+\beta_7)^2$ \\ & $\mu_5 =-\frac{1}{36}(\beta_1+2
        \beta_2-\beta_3+\beta_4+2 \beta_5+\beta_6-2 \beta_7)^2$ \\ & $\mu_6 =
        -\frac{1}{36}(\beta_1-\beta_2+2 \beta_3-2 \beta_4-\beta_5+\beta_6-2 \beta_7)^2$
        \\ & $\mu_7 = -\frac{1}{36}(\beta_1-\beta_2+2 \beta_3+\beta_4+2 \beta_5-2
        \beta_6+\beta_7)^2$ \\ \hline $-\frac{1}{36}(\beta_1 + \beta_2 +\beta_3 -
        \beta_4+\beta_5 + \beta_6 +\beta_7)^2$ & $\mu_1 = -\frac{1}{36}(2 \beta_1+2
        \beta_2+2 \beta_3+\beta_4-\beta_5-\beta_6-\beta_7)^2$ \\ & $\mu_2 =
        -\frac{1}{36}(2 \beta_1-\beta_2-\beta_3-2 \beta_4+2 \beta_5-\beta_6-\beta_7)^2$
        \\ & $\mu_3 = -\frac{1}{36}(2 \beta_1-\beta_2-\beta_3+\beta_4-\beta_5+2
        \beta_6+2 \beta_7)^2$ \\ & $\mu_4 = -\frac{1}{36}(-\beta_1+2 \beta_2-\beta_3-2
        \beta_4-\beta_5+2 \beta_6-\beta_7)^2$ \\ & $\mu_5 = -\frac{1}{36}(-\beta_1+2
        \beta_2-\beta_3+\beta_4+2 \beta_5-\beta_6+2 \beta_7)^2$ \\ & $\mu_6 =
        -\frac{1}{36}(-\beta_1-\beta_2+2 \beta_3-2 \beta_4-\beta_5-\beta_6+2 \beta_7)^2$
        \\ & $\mu_7 = -\frac{1}{36}(-\beta_1-\beta_2+2 \beta_3+\beta_4+2 \beta_5+2
        \beta_6-\beta_7)^2$ \\ \hline
      \end{tabular}} \caption{Metrics associated to different eigenvalues of
      $\Theta$ for the 7-parameter ansatz}
  \end{table}

\end{document}